\documentclass[twocolumn,
superscriptaddress,
prl,
showpacs,
preprintnumbers,
amsmath,amssymb]{revtex4}

\usepackage{graphicx,epsf,epsfig,ulem}
\usepackage{epsfig}
\usepackage{bm}
\usepackage{subfigure}
\usepackage{color}

\begin{document}

\title{Gate induced g-factor control and dimensional transition for donors in multi-valley semiconductors}

\author{Rajib Rahman}
\affiliation{Network for Computational Nanotechnology, Purdue University, West Lafayette, IN 47907, USA}

\author{Seung H. Park}
\affiliation{Network for Computational Nanotechnology, Purdue University, West Lafayette, IN 47907, USA}


\author{Timothy B. Boykin}
\affiliation{University of Alabama, Hunstville, AL 47907, USA}

\author{Gerhard Klimeck}
\affiliation{Network for Computational Nanotechnology, Purdue University, West Lafayette, IN 47907, USA}
\affiliation{Jet Propulsion Laboratory, California Institute of Technology, Pasadena, CA 91109, USA}

\author{Sven Rogge}
\affiliation{Kavli Institute of Nanoscience, Delft University of Technology, Delft, The Netherlands}

\author{Lloyd C. L. Hollenberg}
\affiliation{Center for Quantum Computer Technology, School of Physics, University of Melbourne, VIC 3010, Australia}

\date{\today}

\begin{abstract}
The dependence of the g-factors of semiconductor donors on applied
electric and magnetic fields is of immense importance in spin based
quantum computation and in semiconductor spintronics. The donor
g-factor Stark shift is sensitive to the orientation of the electric
and magnetic fields and strongly influenced by the band-structure
and spin-orbit interactions of the host. Using a multi-million atom
tight-binding framework the spin-orbit Stark parameters are computed
for donors in multi-valley semiconductors, 
silicon and germanium. 
Comparison with limited experimental data
shows good agreement for a donor in silicon. Results for gate induced transition from 3D to
2D wave function confinement show that the corresponding g-factor
shift in Si is experimentally observable.
\end{abstract}


\maketitle Understanding the behavior of single donor electron bound
states under mesoscopic electric and magnetic fields is a
fundamental issue critical to current miniaturization of
semiconductor devices {\cite{Rogge.NaturePhysics.2008}} 
and to the development of new quantum
technologies {\cite{Kane.nature.1998, Vrijen.pra.2000,
Hollenberg.prb.2004}}. It is only very recently that convergence
between experiment and theory has occurred for the electric gate
control of the orbital states {\cite{Rogge.NaturePhysics.2008}} and
electron-nuclear hyperfine interaction {\cite{Bradbury.prl.2006,
Rahman.prl.2007}} for a donor in silicon. However, the Stark shift of the donor spin-orbit
interaction, which is central to understanding the precise spin
properties in combined electric and magnetic fields, is just beginning to be
understood {\cite{De.prl.2009}}. We report the first atomistic treatment of the
donor spin-orbit interaction in multi-valley semiconductors in gated environments, and show non-trivial
agreement with experiment where available. We calculate the donor g-factor
shift for the transition from 3D Coulomb to 2D interface confinement
and show that the effect is experimentally observable.

\begin{figure}[htbp]
\center\epsfxsize=3.0in\epsfbox{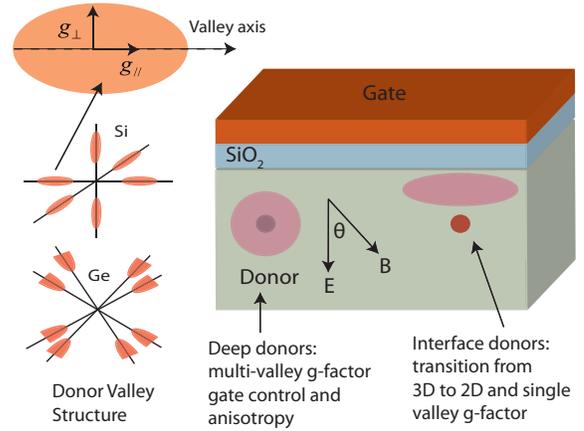}
\caption{Wave function and g-factor engineering of donor and
interface confined electrons by EM fields in Si and Ge
semiconductors (CB equivalent energy surfaces shown left).}
\vspace{-0.5cm}
\end{figure}

Wave function engineering of donor spins is a basic ingredient of
several quantum computing schemes {\cite{Kane.nature.1998,
Vrijen.pra.2000, Hollenberg.prb.2004}}, and may also help realize
novel devices based on spin degrees of freedom. In one method, an
applied E-field deforms the donor wave function and modifies its
orbital angular momentum, which in turn can modify its spin
properties through the spin-orbit interaction. This spin-orbit Stark
effect is manifested by an E-field dependence of the effective
g-factor, and can be probed by ESR experiments
\cite{Bradbury.prl.2006}. However, as we are dealing with donor
levels in the solid-state, generally with complicated multi-valley
orbital-spin effects, the physical origins of this phenomenon is at
present not well understood. In semiconductors with significant
spin-orbit interaction, this technique can even provide a way to
rotate spins by electrical modulation of the g-tensor, and was
demonstrated in GaAs quantum dots {\cite{Nowack.science.2007}} and
in GaAs/Al${}_{x}$Ga${}_{1-x}$As heterostructures
{\cite{Kato.science.2003, Salis.nature.2001}}. While the spin-orbit
interaction in Si is relatively small, in a quantum computer
application such effects can lead to qubit errors at the threshold
level and need to be characterized and understood.

In this letter we report the first investigation of the
Stark shift of the donor g-factor in two multi-valley
semiconductors with varying degrees of spin-orbit interaction. We
quantify the way in which the E-field removes the
isotropy of the donor g-tensor components, resulting in an
anisotropic Zeeman interaction. We also compare our g-factor Stark
shift for P donors in Si against the only available measured value
\cite{Bradbury.prl.2006}, and report corresponding parameters for Ge
host under different orientations of E and B fields, as a
guide for future experiments. Finally, we investigate g-factors of
donors close to an oxide semiconductor interface, and study the
g-factor variation as the electron undergoes a symmetry transition
from 3D Coulomb to 2D interfacial confinement
{\cite{Rogge.NaturePhysics.2008}}. This transition is central to
proposals for donor-gate-confined interfacial transport and qubits
in Si \cite{Skinner.prl.2003,Calderon.prl.2006}.

Engineering the magnetic field response in semiconductors
typically involves compound structures such as
Al${}_x$Ga${}_{1-x}$As or Si${}_x$Ge${}_{1-x}$ with a spatially 
varying material composition. Since the two materials, Al
and Ga in Al${}_x$Ga${}_{1-x}$As for example, have different
g-factors, the effective g-factor of an electronic wave function can
be controlled by pulling the wavefunction from an Al-rich part of
the device to the Ga-rich part by means of an E-field
{\cite{Vrijen.pra.2000}}. The direct dependence of the g-factor on
the field, however, has been largely ignored in literature except
for Refs {\cite{Salis.nature.2001}}, where the g-tensor modulation
resonance was used in Al${}_x$Ga${}_{1-x}$As
hetero-structures to control spin coherence electrically. In Ref
{\cite{Flatte.prl.2006}}, an all electrical control of the spin
of a Mn hole in GaAs was investigated. A large anisotropic Zeeman
splitting has been reported for acceptor levels in SiGeSi quantum wells {\cite{Winkler.prl.2006}} and also for single quantum states of nanoparticles {\cite{Petta.prl.2002}}. 
Past ESR experiments {\cite{Roth.physrev.1960, Wilson.physrev.1961}} 
have investigated the effect of uniaxial strain on donor g-factors in Si
and Ge, while a recent work demonstrated the gate control of spin-orbit interaction in a GaAs/AlGaAs quantum well \cite{gossard.arxiv.2009}.

In this work, we employed an atomistic tight-binding (TB) theory with a 20 orbital
$sp^3d^5s*$ basis per atom including nearest neighbor and spin-orbit
(SO) interactions. The total Hamiltonian of the host and the donor
under an applied E-field can be expressed as,
\begin{equation} \label{eq:hamiltonian}
H = H_0 - \frac{\hbar}{4m_{0}^{2}c^{2}}\vec{\sigma} \cdot \vec{p}
\times \vec{\nabla} V_0  + U_{\rm donor} (r) + e \vec{E} \cdot
\vec{r}.
\end{equation}
The first term represents the host semiconductor, the second term
the SO interaction of the host due to the crystal potential $V_{0}$,
the third and the fourth represent the donor potential and the
applied E-field. 
The semi-empirical TB parameters for Si and Ge {\cite{Boykin.prb.2004, Klimeck.cmes.2002}} used here have been well established in literature. 
The SO interaction of the host was represented as a matrix element between
the p orbitals of the same atom after Chadi {\cite{Chadi.prb.1977}},
and has been shown to cause energy splitting between the split-off
hole (SH) band and the degenerate manifold  of the light (LH) and heavy (HH) hole
bands. This representation includes both the
Rashba and Dresselhaus terms inherently, as opposed to the k$\cdot$p
method where the two are separately expressed. The donors are
represented by a Coulomb potential screened by the dielectric
constant of the host. The potential at the donor site $U_0$ was
adjusted to obtain the ground state (GS) binding energy
{\cite{Shaikh.encyclopedia.2008}} taking into account the
valley-orbit interaction in multi-valley semiconductors
{\cite{Kohn.physrev.1955}}. 
The total Hamiltonian was solved by a parallel
Block Lanczos algorithm to obtain the relevant donor states. A typical
simulation involved about 3 million atoms, and requires about 5 hours on 40 processors. 
The Zeeman Hamiltonian was then evaluated perturbatively, using the matrix
elements ${H_{Z}}_{ij}={\langle \Psi_{i}(\vec{r}, \vec{E})  |
(\vec{L}+2\vec{S}) \cdot{} \vec{B} | \Psi_{j}(\vec{r}, \vec{E})
\rangle}$, where $i,j$ represent $\uparrow$ and $\downarrow$ spins
of a donor state. The small B-field (1T) used throughout this work
justified the inclusion of linear B-field dependencies only in the
Zeeman Hamiltonian. The g-factor was then evaluated using the lowest
spin states ($\epsilon$),
$g(\vec{E},\vec{B})=(\epsilon_{\uparrow}-\epsilon_{\downarrow})/\mu_{B}|\vec{B}|$.

This TB model has been previously used to investigate the
Stark shift of the hyperfine constant for a P donor in Si
{\cite{Rahman.prl.2007}} in good agreement with ESR measurements
{\cite{Bradbury.prl.2006}} and momentum space methods
{\cite{Wellard.prb.2005}}. It has also been successfully applied to interpret
orbital Stark shift measurements  on single As donors in Si FinFETS {\cite{Rogge.NaturePhysics.2008}}.


The g-factor for a donor ground state in a multi-valley
semiconductor is influenced by two main factors. Within a single
valley, the g-factor of electrons moving parallel to the valley axis
($g_{||}$) is different from the g-factor for perpendicular motion
($g_{\vdash}$), assuming the semiconductor has non-spherical energy
surfaces. This anisotropy may be affected further by external
perturbations such as strain or E-fields, which may cause higher
lying conduction bands (CB) to admix with the lowest CB
{\cite{Wilson.physrev.1961}}. Secondly, the donor ground states in
Si and Ge have an equal admixture of all the valleys due to the
valley-orbit interaction {\cite{Kohn.physrev.1955}}, resulting in an
isotropic effective g-factor. Since an E-field removes the
equivalency of the valleys, the effective g-factor becomes
anisotropic depending on the contribution of the different valleys
to the quantum state. The first effect is termed as the
single-valley effect while the second the valley repopulation effect
{\cite{Roth.physrev.1960}}. In a tight-binding description, it is
not necessary to single out different valley contributions since the
full band structure is considered in the formalism. Hence both effects are captured in the
resulting g-factor. 

\vspace{-0.5cm}
\begin{table} [!htb] \caption{Comparison of the
quadratic g-factor Stark shift coefficient for donors in Si and Ge
under different orientation of electric and magnetic fields.}
 \label{tb:tablename}

{\scriptsize
\begin{tabular}{ | c | c | c | c | c | c | c | c }
\hline {Donor} & {Valence band} & {Binding} & {E-field} & {B-field}
& \multicolumn{2}{c|} {$\eta_2$ [$10^{-3}\mu {\rm m}^{2}$/${\rm
V}^{2}$] }
\\

\cline {6-7}

(valleys) & {splitting} & {energy} & & & {Theory} & {Expt
{\cite{Bradbury.prl.2006}}}
\\

{[direction]} & {[eV]} & {[meV]} &  &  &  &
\\

\hline
Si:P & 0.044 & -45.6 & [010] & $B ||$ & -0.012 & -0.01 \\
\cline {5-6}

(6) [100] & & & & $B_\perp$ & 0.014&  (Si:Sb)\\

\hline
Ge:P &  0.29 & -12.8 & [010] & $B ||$ & -4.8 &  - \\
\cline {5-6}

(4) [111] & & & & $B_\perp$ & -4.8 &  \\
\cline {4-6}

& & & [111] & $B||$ & 143.8 &  \\

\cline {5-6}
& & & & $B_\perp$ & -80.1 &  \\



\hline
\end{tabular}
}
\label{ta:1}
\end{table}


Table \ref{ta:1} compares the spin-orbit properties of
donors in Si and Ge. The SO interaction is stronger in Ge than in Si, as shown by the
spin-orbit interaction whose strength is shown by the energy
splitting of the split-off valence band from the degenerate light
and heavy hole bands at the gamma point of its band structure. Si
and Ge are multi-valley semiconductors with valleys located along
[100] and [111] crystal axes respectively.

\begin{figure}[htbp]
\center\epsfxsize=3.4in\epsfbox{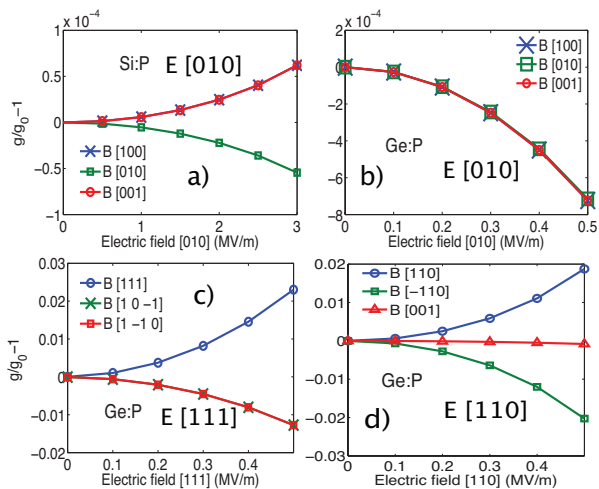} \caption{
Relative change in the the donor g-factor in Si and Ge as a function of
E-field strength (in applied direction) and for parallel and
perpendicular magnetic field orientation. a) Si:P under [010] E-filed , b) Ge:P under [010] E-field, c)
Ge:P under [111] E-field, and d) Ge:P under  [110] E-field. The g-factor shift is sensitive to the relative orientation of the E and B fields with respect to the valley axis.} 
\label{fi:2}
\vspace{-0.5cm}
\end{figure}


Fig \ref{fi:2} shows that the g-factor of donors primarily varies
quadratically with the E-field.
The g-factor shifts are affected by the relative angles between the E-field, the
valley axis, and the  B-field. Fig \ref{fi:2}a shows the g-factor of a P
donor in Si subjected to [010] E-fields, while \ref{fi:2} b, c, and d are for a P donor in Ge under various orientations of the E-field. 
In Si the [010] E-field (2a) removes the equivalency of the six valleys, introducing the valley-repopulation effect in the donor wavefunction. This results in two different parabolas for the g-factor shifts, one for B parallel to E and the other for perpendicular orientation. 
The [010] directed E-field cannot remove the
equivalency of the [111] valleys in Ge, and both parallel and
perpendicular B-fields produce the same g-factor shifts (\ref{fi:2}b). 
However, when the field is directed along the [111] valley
axis as in 2c, we obtain the split
g-factor parabolas (\ref{fi:2}b) similar to Fig 2a. 
In the absence of an E-field, the Zeeman effect of the donor ground state is
isotropic as shown by the convergence of the two parabolas at $\textrm{E}=0$ in both Figs \ref{fi:2}a and \ref{fi:2}c.





The results of Fig \ref{fi:2} are fitted to a quadratic equation
$g(\textrm{E})/g(0)-1 = \eta_2 \textrm{E}^2$, where $\eta_2$ 
is the quadratic Stark coefficient. Values of $\eta_2$ are shown in Table \ref{ta:1} for a few different E and B field orientations.
For donors close to interfaces, there can be some linear Stark effect as well, which we discuss briefly in Fig 4. The quadratic coefficient of Si:P is of the order of
$10^{-5}$ ${\mu}{\rm m}^{2}/{\rm V}^{2}$, and compares well in
magnitude with the only available experimental data on Si:Sb. Order
of magnitude comparison of $\eta_2$ for Si and Ge shows
that the spin-orbit Stark effect is stronger in Ge than in
Si. The Zeeman anisotropy is also stronger for donors in Ge, where the quadratic coefficients can differ by an order of
magnitude between parallel and perpendicular B-fields (Table 1).
The direction of the E-field relative to the valley-axes also affects the
strength of the Zeeman interaction. 
This is shown by comparing the $B_{||}$ results of Ge:P for [010] and [111]
directed E-fields. The quadratic Stark coefficients in this case
differ by two orders of magnitude. 
Our results show very good agreement in magnitude of $\eta_2$ for
Si:P with the measured value reported in Ref \cite{Bradbury.prl.2006}, however, there is a discrepancy regarding the relative sign of the g-factor shift in the quadratic regime which we are currently unable to account for. Within the TB framework calculations of the g-factor Stark shift for the single valley GaAs case were also carried out and showed good agreement in magnitude in comparison with the recent k.p results \cite{De.prl.2009}. 



A simple multi-valley picture provides some intuitive explanations of the Stark shifted g-factor based on the valley repopulation effect. 
If $|a_x|^2$ represents the contribution of the $+x$ valley in Si to the donor GS, and $g^{'}_{+x}$ the diagonal g-tensor corresponding to this valley with the x component given by $g_{||}$ while the y and z components given by $g_{\perp}$, then the effective g-tensor of the donor GS is given by $g(E)= \sum_{i=\pm{x},\pm{y},\pm{z}}{|a_i|^2g^{'}_i}$.
Assuming $a_i=a_{-i}$ and $a_x=a_z$ for a [010] E-field, we obtain the effective g-tensor components $g_x$, $g_y$ and $g_z$ as, 
\begin{eqnarray} 
g_x & = & g_z=2(|a_y|^2+|a_x|^2)g_{\perp}+2|a_x|^2g_{||} \\
\label{eq:g1}
g_y & = & 4|a_x|^2g_{\perp}+|a_y|^2g_{||}
\label{eq:g2}
\end{eqnarray}

These equations show that the parallel component of the g-factor has a different response to the electric field as compared to the perpendicular component, verifying the split g-factor curves of Fig \ref{fi:2}a and \ref{fi:2}d. At $\textrm{E}=0$, each $a_i=1/\sqrt{6}$, and eqs (2) and (3) reduce to $g_x=g_y=g_z=\frac{2}{3}g_{\perp}+\frac{1}{3}g_{||}=g_0$, showing an isotropic effective g-factor. At ionizing E-fields, only the valleys parallel to the field contribute to the state. Setting $a_y=1/\sqrt{2}$ and $a_x=a_z=0$ in (2)  and (3), $g_x=g_{\perp}$ and $g_y=g_{||}$, which helps to probe the single valley g-factors, as shown later in Fig \ref{fi:4}. 
Similar expressions can be derived for a donor in Ge taking into account that the Ge valleys are in [111]. For a more quantitative approach, however, one needs to know also the g-factor variation within a single valley, the precise nature of the wavefunction distortion by the E-field, and the effect of B-fields. The TB approach provides a generalized framework to include all these.


\begin{figure}[htbp]
\center\epsfxsize=3.2in\epsfbox{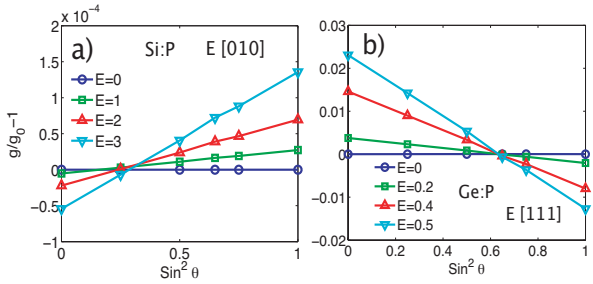} \caption{Anisotropic
Zeeman effect with E-field. $\theta$ represents the angle between
the E and the 1T B field. a) Si:P under [010] E-field with B-field
varying from [010] to [001]. b) Ge:P under [111] E-field, with
B-field varying from [111] to [1-10]. }
\label{fi:3}
\vspace{-0.5cm}\end{figure}

In Fig \ref{fi:3}, we vary the angle $\theta$ between the E and
B-fields from 0 to 90 degrees for a) Si:P under [010] E-field, and
b) Ge:P under [111] E-field. The relative change in g-factor shows a
linear dependence on $\sin^{2}\theta$, consistent with Ref
{\cite{Wilson.physrev.1961}}. The sensitivity of this variation
increases at higher E-fields as shown in Fig \ref{fi:3}a and \ref{fi:3}b. The flat $\textrm{E}=0$
line indicates that the Zeeman effect is isotropic at zero field. Assuming an y-directed E-field in Si such that
the effective g-tensor diagonal components are given by $g_x(E)=g_z(E) \neq g_y(E)$, the linear dependence of $g(E)$ on $\sin^2 \theta$ is shown by,
\begin{equation} \label{eq:gB1}
g(E) \approx g_{y}(E)(1+\frac{{g_{x}(E)}^2-{g_{y}(E)}^2}{2{g_{y}(E)}^2} \sin^2 \theta)
\end{equation}
Eq \ref{eq:gB1} can be derived by expanding $g= ({g_{y}(E)}^2 \cos^2 \theta + {g_{x}(E)}^2 \sin^2 \theta)^{1/2}$ up to linear terms in $({g_{x}(E)}^2/{g_{y}(E)}^2-1)\sin^2 \theta$.

\begin{figure}[htbp]
\center\epsfxsize=3.4in\epsfbox{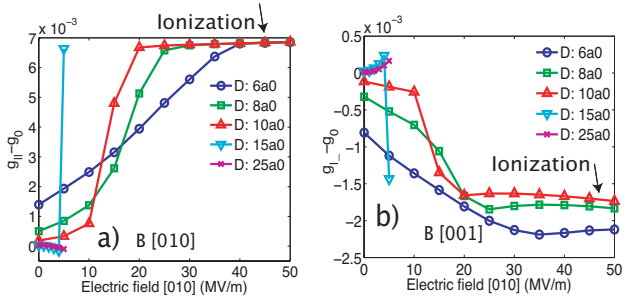} \caption{ Interface
effects on the donor g-factor for Si:P. g-component a) parallel, b)
perpendicular to the field and the interface at various donor
depths. } 
\label{fi:4}
\vspace{-0.3cm}\end{figure}


The confinement transition from 3D Coulomb to an interface 2D system
has recently been observed (in conjunction with the theoretical
approach used here) \cite{Rogge.NaturePhysics.2008}, and is related
to a new control scheme based on dopants close to the Si-SiO$_2$
interface {\cite{Calderon.prl.2006}}. The donor electron can be
adiabatically pulled to the interface by gate voltages
{\cite{Martins.prb.2004, Smit.prb.2004,Wellard.prb.2005}, and
controlled by surface gates. We computed the g-factor of a system
undergoing this confinement transition. Fig \ref{fi:4} shows the components
of the g-factor parallel (\ref{fi:4}a) perpendicular (\ref{fi:4}b) to the E-field for
various donor depths. As the E-field increases, the two Si
conduction band valleys in the direction of the field are lowered in
energy relative to the four valleys perpendicular to the field axis.
The interface state realized at ionizing E-fields has contribution
from these two uni-axial valleys, and hence their $g_{||}$ and
$g_{\perp}$ approach those of the two valleys. Our simulations
indicate $g_{||}-g_{\perp} \approx 8 \times 10^{-3}$, which
compares well in order of magnitude with the measurements of Ref
{\cite{Wilson.physrev.1961}}. This g-factor anisotropy has also been
reported in 2DEGS {\cite{Winkler.prl.2000}}. The transition to the
single valley g components is abrupt if the donor is far away from
the interface ($>$10 nm), and gradual if the donors are closer to
the interface \cite{Smit.prb.2004, Martins.prb.2004,
Wellard.prb.2005}. Proximity to interfaces is also marked by linear
Stark effect since the wave function becomes asymmetric due to sharp
truncation by the surface. Fig \ref{fi:4} shows at small donor depths, the $g_{||}$ and
$g_{\perp}$ also exhibit a linear field dependence. In this regime, the linear Stark coefficient
exceeds the quadratic coefficient. A similar effect was observed for
the hyperfine Stark effect {\cite{Rahman.prl.2007}}.

In conclusion, we have for the first time applied atomistic
techniques to understand and predict the E-field response of donor
g-factors in multi-valley (Si, Ge) semiconductors
with different degrees of spin-orbit interaction. The E-field
induces a Zeeman anisotropy that varies with the relative angle
between the E and the B fields. The strength of the Stark shift is
also dependent on the direction of the E-field relative to the valley
axis. The computed Stark shift coefficient of Si:P compares well in
magnitude with the only available measured value for donors in Si.
The donor g-factor Stark shift was also computed for the 3D to 2D
confinement transition, suggesting that the effect is accessible to
experiments.


This work was supported by the Australian Research Council, NSA and
ARO  (contract number W911NF-04-1-0290). Part of the development of
NEMO-3D was performed at JPL, Caltech under a contract with NASA.
NCN/nanohub.org computer resources were used.

Electronic address: rrahman@purdue.edu

\end{document}